\documentclass[aps,prd,superscriptaddress,twocolumn]{revtex4-1}
\usepackage{graphicx,url,amssymb,amsmath,longtable,rotating,color,units,wasysym,subfigure,epsfig}

\begin{document}

\title{Improved Upper Limits on the Stochastic Gravitational-Wave Background from 2009-2010 LIGO and Virgo Data}



\author{%
J.~Aasi$^{1}$,
B.~P.~Abbott$^{1}$,
R.~Abbott$^{1}$,
T.~Abbott$^{2}$,
M.~R.~Abernathy$^{1}$,
T.~Accadia$^{3}$,
F.~Acernese$^{4,5}$,
K.~Ackley$^{6}$,
C.~Adams$^{7}$,
T.~Adams$^{8}$
% ...
}\noaffiliation

\affiliation {LIGO - California Institute of Technology, Pasadena, CA 91125, USA }
\affiliation {Louisiana State University, Baton Rouge, LA 70803, USA }
\affiliation {Laboratoire d'Annecy-le-Vieux de Physique des Particules (LAPP), Universit\'e de Savoie, CNRS/IN2P3, F-74941 Annecy-le-Vieux, France }
\affiliation {INFN, Sezione di Napoli, Complesso Universitario di Monte S.Angelo, I-80126 Napoli, Italy }
\affiliation {Universit\`a di Salerno, Fisciano, I-84084 Salerno, Italy }
\affiliation {University of Florida, Gainesville, FL 32611, USA }
\affiliation {LIGO - Livingston Observatory, Livingston, LA 70754, USA }
\affiliation {Cardiff University, Cardiff, CF24 3AA, United Kingdom }
\affiliation {Albert-Einstein-Institut, Max-Planck-Institut f\"ur Gravitationsphysik, D-30167 Hannover, Germany }
\affiliation {Nikhef, Science Park, 1098 XG Amsterdam, The Netherlands }
\affiliation {LIGO - Massachusetts Institute of Technology, Cambridge, MA 02139, USA }
\affiliation {Instituto Nacional de Pesquisas Espaciais, 12227-010 - S\~{a}o Jos\'{e} dos Campos, SP, Brazil }
\affiliation {Inter-University Centre for Astronomy and Astrophysics, Pune - 411007, India }
\affiliation {Tata Institute for Fundamental Research, Mumbai 400005, India }
\affiliation {Syracuse University, Syracuse, NY 13244, USA }
\affiliation {University of Wisconsin--Milwaukee, Milwaukee, WI 53201, USA }
\affiliation {Leibniz Universit\"at Hannover, D-30167 Hannover, Germany }
\affiliation {INFN, Sezione di Pisa, I-56127 Pisa, Italy }
\affiliation {Universit\`a di Siena, I-53100 Siena, Italy }
\affiliation {Stanford University, Stanford, CA 94305, USA }
\affiliation {The University of Mississippi, University, MS 38677, USA }
\affiliation {California State University Fullerton, Fullerton, CA 92831, USA }
\affiliation {INFN, Sezione di Roma, I-00185 Roma, Italy }
\affiliation {University of Birmingham, Birmingham, B15 2TT, United Kingdom }
\affiliation {Albert-Einstein-Institut, Max-Planck-Institut f\"ur Gravitationsphysik, D-14476 Golm, Germany }
\affiliation {Montana State University, Bozeman, MT 59717, USA }
\affiliation {European Gravitational Observatory (EGO), I-56021 Cascina, Pisa, Italy }
\affiliation {LIGO - Hanford Observatory, Richland, WA 99352, USA }
\affiliation {SUPA, University of Glasgow, Glasgow, G12 8QQ, United Kingdom }
\affiliation {APC, AstroParticule et Cosmologie, Universit\'e Paris Diderot, CNRS/IN2P3, CEA/Irfu, Observatoire de Paris, Sorbonne Paris Cit\'e, 10, rue Alice Domon et L\'eonie Duquet, F-75205 Paris Cedex 13, France }
\affiliation {Columbia University, New York, NY 10027, USA }
\affiliation {Universit\`a di Pisa, I-56127 Pisa, Italy }
\affiliation {CAMK-PAN, 00-716 Warsaw, Poland }
\affiliation {Astronomical Observatory Warsaw University, 00-478 Warsaw, Poland }
\affiliation {INFN, Sezione di Genova, I-16146 Genova, Italy }
\affiliation {Universit\`a degli Studi di Genova, I-16146 Genova, Italy }
\affiliation {San Jose State University, San Jose, CA 95192, USA }
\affiliation {Faculty of Physics, Lomonosov Moscow State University, Moscow 119991, Russia }
\affiliation {LAL, Universit\'e Paris-Sud, IN2P3/CNRS, F-91898 Orsay, France }
\affiliation {NASA/Goddard Space Flight Center, Greenbelt, MD 20771, USA }
\affiliation {University of Western Australia, Crawley, WA 6009, Australia }
\affiliation {Department of Astrophysics/IMAPP, Radboud University Nijmegen, P.O. Box 9010, 6500 GL Nijmegen, The Netherlands }
\affiliation {Universit\'e Nice-Sophia-Antipolis, CNRS, Observatoire de la C\^ote d'Azur, F-06304 Nice, France }
\affiliation {Institut de Physique de Rennes, CNRS, Universit\'e de Rennes 1, F-35042 Rennes, France }
\affiliation {Laboratoire des Mat\'eriaux Avanc\'es (LMA), IN2P3/CNRS, Universit\'e de Lyon, F-69622 Villeurbanne, Lyon, France }
\affiliation {Washington State University, Pullman, WA 99164, USA }
\affiliation {INFN, Sezione di Perugia, I-06123 Perugia, Italy }
\affiliation {INFN, Sezione di Firenze, I-50019 Sesto Fiorentino, Firenze, Italy }
\affiliation {Universit\`a degli Studi di Urbino 'Carlo Bo', I-61029 Urbino, Italy }
\affiliation {University of Oregon, Eugene, OR 97403, USA }
\affiliation {Laboratoire Kastler Brossel, ENS, CNRS, UPMC, Universit\'e Pierre et Marie Curie, F-75005 Paris, France }
\affiliation {VU University Amsterdam, 1081 HV Amsterdam, The Netherlands }
\affiliation {University of Maryland, College Park, MD 20742, USA }
\affiliation {University of Massachusetts - Amherst, Amherst, MA 01003, USA }
\affiliation {Universitat de les Illes Balears, E-07122 Palma de Mallorca, Spain }
\affiliation {Universit\`a di Napoli 'Federico II', Complesso Universitario di Monte S.Angelo, I-80126 Napoli, Italy }
\affiliation {Canadian Institute for Theoretical Astrophysics, University of Toronto, Toronto, Ontario, M5S 3H8, Canada }
\affiliation {Tsinghua University, Beijing 100084, China }
\affiliation {University of Michigan, Ann Arbor, MI 48109, USA }
\affiliation {Rochester Institute of Technology, Rochester, NY 14623, USA }
\affiliation {INFN, Sezione di Roma Tor Vergata, I-00133 Roma, Italy }
\affiliation {National Tsing Hua University, Hsinchu Taiwan 300 }
\affiliation {Charles Sturt University, Wagga Wagga, NSW 2678, Australia }
\affiliation {Caltech-CaRT, Pasadena, CA 91125, USA }
\affiliation {Pusan National University, Busan 609-735, Korea }
\affiliation {Australian National University, Canberra, ACT 0200, Australia }
\affiliation {Carleton College, Northfield, MN 55057, USA }
\affiliation {INFN, Gran Sasso Science Institute, I-67100 L'Aquila, Italy }
\affiliation {Universit\`a di Roma Tor Vergata, I-00133 Roma, Italy }
\affiliation {Universit\`a di Roma 'La Sapienza', I-00185 Roma, Italy }
\affiliation {University of Brussels, Brussels 1050 Belgium }
\affiliation {Sonoma State University, Rohnert Park, CA 94928, USA }
\affiliation {Embry-Riddle Aeronautical University, Prescott, AZ 86301, USA }
\affiliation {The George Washington University, Washington, DC 20052, USA }
\affiliation {University of Cambridge, Cambridge, CB2 1TN, United Kingdom }
\affiliation {Northwestern University, Evanston, IL 60208, USA }
\affiliation {University of Minnesota, Minneapolis, MN 55455, USA }
\affiliation {The University of Texas at Brownsville, Brownsville, TX 78520, USA }
\affiliation {The University of Sheffield, Sheffield S10 2TN, United Kingdom }
\affiliation {Wigner RCP, RMKI, H-1121 Budapest, Konkoly Thege Mikl\'os\'ut 29-33, Hungary }
\affiliation {University of Sannio at Benevento, I-82100 Benevento, Italy }
\affiliation {INFN, Gruppo Collegato di Trento, I-38050 Povo, Trento, Italy }
\affiliation {Universit\`a di Trento, I-38050 Povo, Trento, Italy }
\affiliation {Montclair State University, Montclair, NJ 07043, USA }
\affiliation {The Pennsylvania State University, University Park, PA 16802, USA }
\affiliation {MTA E\"otv\"os University, `Lendulet' A. R. G., Budapest 1117, Hungary }
\affiliation {Universit\`a di Perugia, I-06123 Perugia, Italy }
\affiliation {Rutherford Appleton Laboratory, HSIC, Chilton, Didcot, Oxon, OX11 0QX, United Kingdom }
\affiliation {Perimeter Institute for Theoretical Physics, Ontario, N2L 2Y5, Canada }
\affiliation {American University, Washington, DC 20016, USA }
\affiliation {University of Adelaide, Adelaide, SA 5005, Australia }
\affiliation {Raman Research Institute, Bangalore, Karnataka 560080, India }
\affiliation {Korea Institute of Science and Technology Information, Daejeon 305-806, Korea }
\affiliation {Bia{\l }ystok University, 15-424 Bia{\l }ystok, Poland }
\affiliation {University of Southampton, Southampton, SO17 1BJ, United Kingdom }
\affiliation {IISER-TVM, CET Campus, Trivandrum Kerala 695016, India }
\affiliation {Institute of Applied Physics, Nizhny Novgorod, 603950, Russia }
\affiliation {Seoul National University, Seoul 151-742, Korea }
\affiliation {Hanyang University, Seoul 133-791, Korea }
\affiliation {IM-PAN, 00-956 Warsaw, Poland }
\affiliation {NCBJ, 05-400\'Swierk-Otwock, Poland }
\affiliation {Institute for Plasma Research, Bhat, Gandhinagar 382428, India }
\affiliation {The University of Melbourne, Parkville, VIC 3010, Australia }
\affiliation {INFN, Sezione di Padova, I-35131 Padova, Italy }
\affiliation {Monash University, Victoria 3800, Australia }
\affiliation {SUPA, University of Strathclyde, Glasgow, G1 1XQ, United Kingdom }
\affiliation {ESPCI, CNRS, F-75005 Paris, France }
\affiliation {Argentinian Gravitational Wave Group, Cordoba Cordoba 5000, Argentina }
\affiliation {Universit\`a di Camerino, Dipartimento di Fisica, I-62032 Camerino, Italy }
\affiliation {The University of Texas at Austin, Austin, TX 78712, USA }
\affiliation {Southern University and A\&M College, Baton Rouge, LA 70813, USA }
\affiliation {College of William and Mary, Williamsburg, VA 23187, USA }
\affiliation {IISER-Kolkata, Mohanpur, West Bengal 741252, India }
\affiliation {National Institute for Mathematical Sciences, Daejeon 305-390, Korea }
\affiliation {Hobart and William Smith Colleges, Geneva, NY 14456, USA }
\affiliation {RRCAT, Indore MP 452013, India }
\affiliation {SUPA, University of the West of Scotland, Paisley, PA1 2BE, United Kingdom }
\affiliation {Institute of Astronomy, 65-265 Zielona G\'ora, Poland }
\affiliation {Indian Institute of Technology, Gandhinagar Ahmedabad Gujarat 382424, India }
\affiliation {Instituto de F\'\i sica Te\'orica, Univ. Estadual Paulista/International Center for Theoretical Physics-South American Institue for Research, S\~ao Paulo SP 01140-070, Brazil }
\affiliation {Andrews University, Berrien Springs, MI 49104, USA }
\affiliation {Trinity University, San Antonio, TX 78212, USA }
\affiliation {University of Washington, Seattle, WA 98195, USA }
\affiliation {Southeastern Louisiana University, Hammond, LA 70402, USA }
\affiliation {Abilene Christian University, Abilene, TX 79699, USA }


\begin{abstract}
Gravitational waves from a variety of sources are predicted to superpose to create a stochastic background.  This background is expected to contain unique information from throughout the history of the universe that is unavailable through standard electromagnetic observations, making its study of fundamental importance to understanding the evolution of the universe.  We carry out a search for the stochastic background with the latest data from LIGO and Virgo.  Consistent with predictions from most stochastic gravitational-wave background models, the data display no evidence of a stochastic gravitational-wave signal.  Assuming a gravitational-wave spectrum of ${\Omega}_{\rm GW}(f)={\Omega}_{\alpha}\left({f/f_{\rm{ref}}}\right)^{\alpha}$, we place 95\% confidence level upper limits on the energy density of the background in each of four frequency bands spanning 41.5--1726~Hz.  
In the frequency band of 41.5--169.25~Hz for a spectral index of  $\alpha=0$, we constrain the energy density of the stochastic background to be $\Omega_{\rm GW}(f)<5.6\times10^{-6}$.   For the 600--1000~Hz band, $\Omega_{\rm GW}(f)<0.14(f/900~{\rm Hz})^3$, a factor of 2.5 lower than the best previously reported upper limits.  We find $\Omega_{\rm GW}(f)<1.8\times10^{-4}$ using a spectral index of zero for 170--600~Hz and $\Omega_{\rm GW}(f)<1.0(f/1300~{\rm Hz})^3$ for 1000--1726~Hz, bands in which no previous direct limits have been placed.  The limits in these four bands are the lowest direct measurements to date on the stochastic background.  We discuss the implications of these results in light of the recent claim by the BICEP2 experiment of the possible evidence for inflationary gravitational waves.
\end{abstract}

\maketitle

\emph{Introduction.}---The stochastic gravitational-wave background (SGWB) has great potential to be a rich area of study since it is expected to include contributions from a superposition of astrophysical and/or cosmological sources.  Astrophysical contributions to the background might very well dominate in the LIGO/Virgo frequency band.  These contributions may include compact binary coalescences \citep{2013MNRAS.431..882Z,PhysRevD.84.124037,PhysRevD.85.104024,PhysRevD.84.084004,2011ApJ...739...86Z}, rotating neutron stars \citep{PhysRevD.87.063004,2012PhRvD..86j4007R,2011ApJ...729...59Z}, magnetars \citep{2011MNRAS.411.2549M,2006A&A...447....1R,2013PhRvD..87d2002W}, and supernovae \citep{2009MNRAS.398..293M,2010MNRAS.409L.132Z,PhysRevD.72.084001,PhysRevD.73.104024}.  Many mechanisms for generating cosmological contributions to the stochastic background have been postulated as well, such as inflationary models \citep{1994PhRvD..50.1157B,1979JETPL..30..682S,2007PhRvL..99v1301E,2012PhRvD..85b3525B,2012PhRvD..85b3534C,2013arXiv1305.5855L,1997PhRvD..55..435T,2006JCAP...04..010E} and cosmic strings \citep{2005PhRvD..71f3510D,1976JPhA....9.1387K,2002PhLB..536..185S,2007PhRvL..98k1101S}.  
The recent observation of B-mode polarization in the cosmic microwave background claimed by the BICEP2 experiment \citep{PhysRevLett.112.241101}, when using common dust emission models, suggests the presence of gravitational waves produced by primordial vacuum modes amplified by inflation (although the lack of public dust emission maps means BICEP2 could not empirically exclude dust emission as being wholly responsible for the excess B-mode polarization and recent analyses reinforce this \citep{2014arXiv1405.7351F,2014arXiv1405.5857M}).   The energy density of these gravitational waves in the LIGO/Virgo frequency band is several orders of magnitude weaker than typical predictions for astrophysical contributions and six orders of magnitude weaker than what Advanced LIGO \citep{2010CQGra..27h4006H} and Advanced Virgo \citep{Virgo2009} detectors are expected to achieve.  However, non-standard inflationary models \citep{2012PhRvD..85b3525B,2012PhRvD..85b3534C} might surpass even the predicted astrophysical contributions at the LIGO/Virgo frequencies, thereby facilitating detection with Advanced LIGO and Advanced Virgo to which the BICEP2 measurement is not sensitive.  Current alternative theories of inflation, predicting a high-frequency background detectable with Advanced LIGO and Advanced Virgo, remind us that many details of inflation are still unknown, and reality may be more complicated than predicted by simple slow-roll models.  Other cosmological backgrounds, e.g., from cosmic super(strings), may be detectable as well \citep{2007PhRvL..98k1101S}.

The multitude of astrophysical and cosmological sources potentially contributing to a stochastic background offers an opportunity to study many aspects of the universe that are not accessible through standard electromagnetic astrophysical observations \citep{2000PhR...331..283M}.  With the possible observation of a gravitational-wave (GW) imprint on the cosmic microwave background (CMB) \citep{PhysRevLett.112.241101}, we enter an exciting new phase in GW cosmology in which it appears plausible to study the physics of very early times and very high energies.

In this paper we report on a search for the isotropic stochastic background using data gathered in 2009--2010 by LIGO and Virgo.  For the search, we cross-correlated data streams from different detectors to look for correlated stochastic signal.  Most SGWB models predict backgrounds much lower than these data were capable of detecting.  However, this work sets the stage for the Advanced LIGO and Advanced Virgo detectors, which are expected to achieve four orders of magnitude improvement in sensitivity to the GW energy density at 100~Hz and be sensitive to frequencies down to 10~Hz.  Having found no statistically significant evidence of a stochastic gravitational$\mbox{-}$wave signal, we present the best constraints to date on the energy density of the SGWB from LIGO and Virgo.


\emph{Data.}---Previous to this analysis, the best limits on the SGWB from LIGO and Virgo data were obtained using 2005--2007 data \citep{2009Natur.460..990A,2012PhRvD..85l2001A,2011PhRvL.107A1102A}.  For this study, we use data from the LIGO observatories in Hanford, WA (H1) and Livingston Parish, LA (L1)  \citep{2009RPPh...72g6901A} as well as the Virgo observatory in Cascina, Italy (V1) \citep{2012JInst...7.3012A}.  The H2 observatory in Hanford, WA was decommissioned before these data were collected.  LIGO data ran from July 2009-October 2010.  Virgo data spanned July 2009-January 2010 and July 2010-October 2010.


\emph{Method.}---The SGWB energy density spectrum is defined as
\begin{equation}
\Omega_{\rm{GW}}(f)=\frac{f}{\rho_{\rm{c}}}\frac{d\rho_{\rm{GW}}}{df},
\end{equation}
where $f$ is frequency, $\rho_{\rm{c}}$ is the critical (closure) energy density of the universe, and $d\rho_{\rm{GW}}$ is the gravitational radiation energy density contained in the range $f$ to $f+df$ \citep{1999PhRvD..59j2001A}.  For the LIGO and Virgo frequency bands, most theoretical models are characterized by a power law spectrum so we assume the gravitational$\mbox{-}$wave spectrum to be 
\citep{1999PhRvD..59j2001A,2007ApJ...659..918A,2012PhRvD..85l2001A}:
\begin{equation}
{\Omega}_{\rm GW}(f)={\Omega}_{\alpha}\left({\frac{f}{f_{\rm{ref}}}}\right)^{\alpha}.
\label{eq-2}
\end{equation}
Here, $f_{\rm ref}$ is an arbitrary reference frequency (see Table~\ref{tbl-1}).  $\Omega_{\alpha}$ is a constant characterizing the amplitude of the SGWB in a given frequency band.  
Following the precedent of \citep{2009Natur.460..990A,2012PhRvD..85l2001A,2011PhRvL.107A1102A}, we consider two spectral index values: $\alpha=0$ (cosmologically motivated) and $\alpha=3$ (astrophysically motivated).  

We employ a cross$\mbox{-}$correlation method optimized for detecting an isotropic SGWB using pairs of detectors \citep{1999PhRvD..59j2001A}.  This method defines a cross$\mbox{-}$correlation estimator:\newline
\begin{equation}
\hat{Y}=\int_{-\infty}^{\infty}df\int_{-\infty}^{\infty}df'\delta_T(f-f'){\tilde s}^{*}_{1}(f){\tilde s}_{2}(f'){\tilde Q}(f')
\label{eq-3}
\end{equation}
and its variance:
\begin{equation}
\sigma^2_Y{\approx}\frac{T}{2}\int_{0}^{\infty}dfP_1(f)P_2(f)|{{\tilde Q}(f)}|^2,
\end{equation}
where $\delta_T(f-f')$ is the finite$\mbox{-}$time approximation to the Dirac delta function, ${\tilde s}_{1}$ and ${\tilde s}_{2}$ are Fourier transforms of time$\mbox{-}$series strain data from two interferometers, $T$ is the coincident observation time, and $P_1$ and $P_2$ are one$\mbox{-}$sided strain power spectral densities from the two interferometers.  The filter function ${\tilde Q}$ is given by:
\begin{equation}
{\tilde Q}(f)={\lambda}\frac{{\gamma}(f){\Omega}_{\rm GW}(f)H_0^2}{f^{3}P_1(f)P_2(f)},
\end{equation}
where $\lambda$ is a normalization constant chosen such that $\langle \hat{Y}\rangle=\Omega_\alpha$, $\gamma(f)$ is the overlap reduction function arising from the combined antenna patterns of differing detector locations and orientations \citep{PhysRevD.46.5250}, and $H_0$ is the present best estimate of the Hubble constant, $68~{\rm km~s^{-1}~Mpc^{-1}}$ \citep{2013arXiv1303.5062P}.

To combine the measured $\hat{Y}$ for each of the H1L1, H1V1, and L1V1 detector pairs, we follow \citep{1999PhRvD..59j2001A} and average results from detector pairs weighted by their variances.  The optimal estimator is thus given by
\begin{equation}
{\hat{Y}_{\rm tot}=\frac{\sum_l\hat{Y}_l\sigma_l^{-2}}{\sum_l\sigma_l^{-2}}}
\end{equation}
where $l$ sums over detector pairs.  The total variance, $\sigma^2_{\rm tot}$, is 
\begin{equation}
{\sigma_{\rm tot}^{-2}=\sum_l\sigma_l^{-2}.}
\end{equation}


\emph{Analysis.}---Following \citep{2009Natur.460..990A,2012PhRvD..85l2001A,2011PhRvL.107A1102A}, we divide the strain time series data, down$\mbox{-}$sampled to 4096~Hz, into 50\% overlapping 60~s segments that are Hann$\mbox{-}$windowed and high$\mbox{-}$pass filtered with a $6^{\rm th}$ order Butterworth filter with knee frequency 32~Hz. The data are coarse-grained to obtain a frequency resolution of  0.25~Hz.

We include in the analysis only those times when a detector pair has both detectors in a low noise science mode.  Excluded times fall into two different categories.  We exclude data (i) from times when detector operation is unstable and (ii) from times associated with hardware injections, where simulated signals are induced by coherent movement of interferometer mirrors.  These cuts cause $<$2\% reduction in coincident data for each detector pair.  Additionally, we exclude data segments that deviate from the assumption that the power spectra of the detector noise are stationary with time \citep{2009Natur.460..990A}.  Depending on the frequency band, this process excludes up to 4.7\% of data segments.  Combining the above effects, the cuts leave $\sim$117~days of live time for the H1L1 detector pair, $\sim$74~days for H1V1, and $\sim$59~days for L1V1.

Instrumental artifacts can appear in the frequency domain.  We identify high coherence bins using the same method as \citep{2009Natur.460..990A}.  Lines of excess coherence are caused, for example, by power line harmonics and 16~Hz harmonics from H1 and L1 data acquisition systems.  These frequency bins are excluded from the final analysis.  

In order to have an end$\mbox{-}$to$\mbox{-}$end test of the detectors and the analysis pipeline, simulations of a stochastic signal are made in both the hardware and the software (by the addition of a stochastic signal to interferometer data).  
The successful recovery of hardware injections is described in~\citep{Charl2011}.
We successfully recovered a software injection, which had  ${\Omega}_0=1.2\times10^{-4}$ (corresponding to a signal-to-noise ratio of $\approx10$), in all three detector pairs using about one third of the data.

Coherence studies have been made comparing data from magnetometers at 
the LIGO Hanford, LIGO Livingston and Virgo observatories 
\citep{2013PhRvD..87l3009T}. These studies report on the observations of 
correlated magnetic field noise between observatories and its potential coupling to GW detectors. While this may be a concern for future 
generations of detectors with their improved sensitivity, it does not affect the data used in the 
analysis presented in this paper or previous results \citep{2009Natur.460..990A,2012PhRvD..85l2001A,2011PhRvL.107A1102A}.

\begin{figure*} 
\centering
 \begin{tabular}{cc}
    \psfig{file=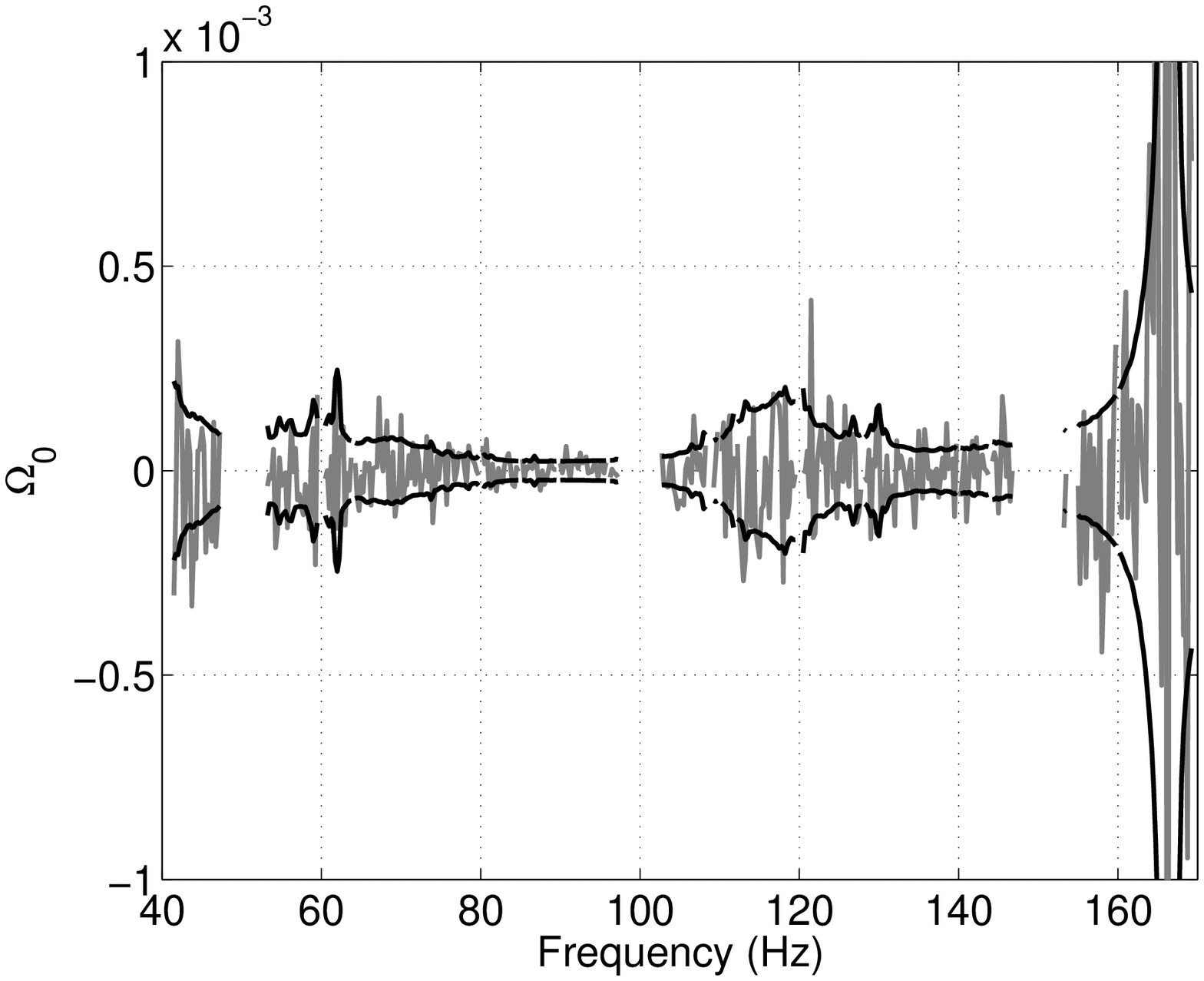, width=0.48\textwidth} &
    \psfig{file=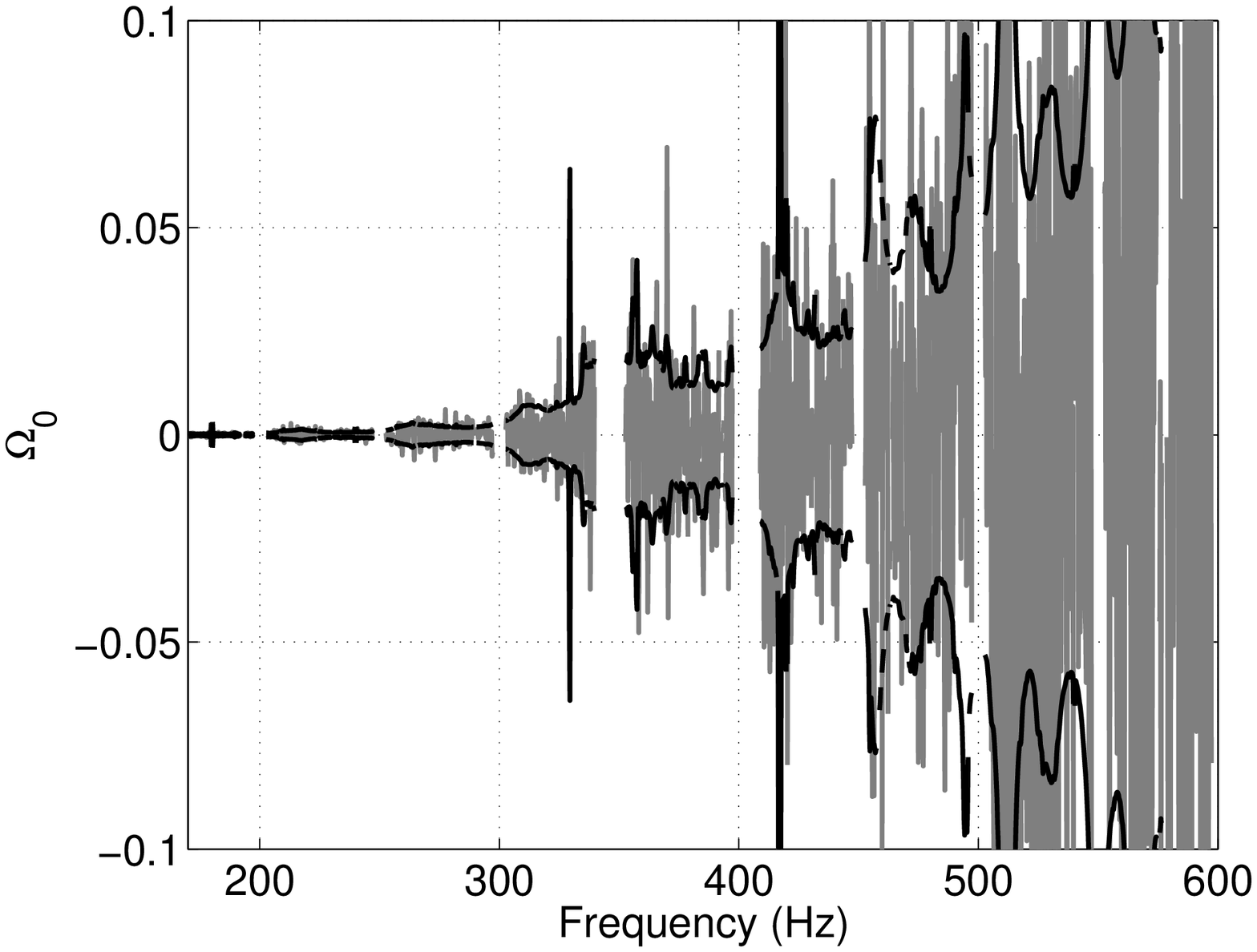, width=0.48\textwidth} \\
    \psfig{file=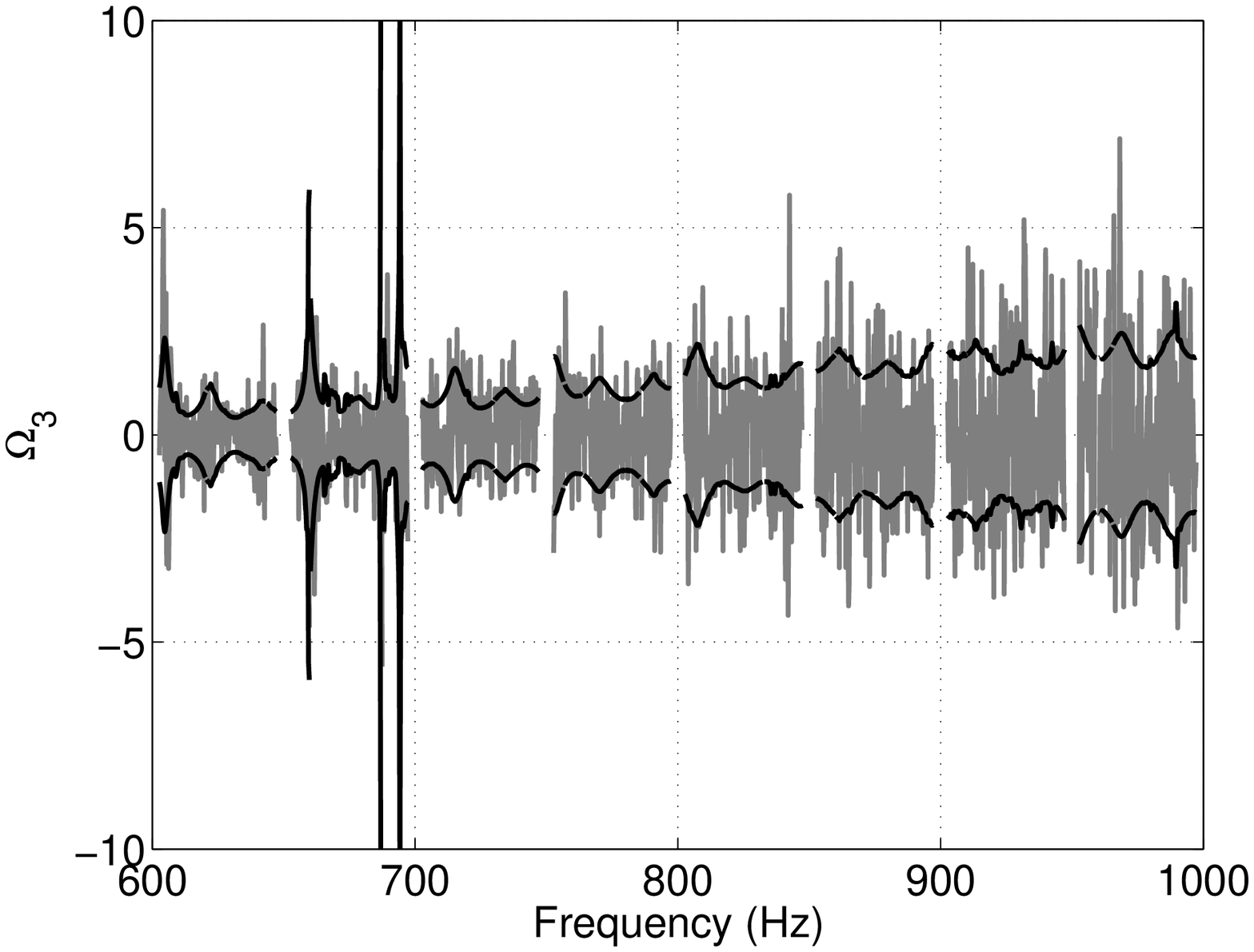, width=0.48\textwidth} &
    \psfig{file=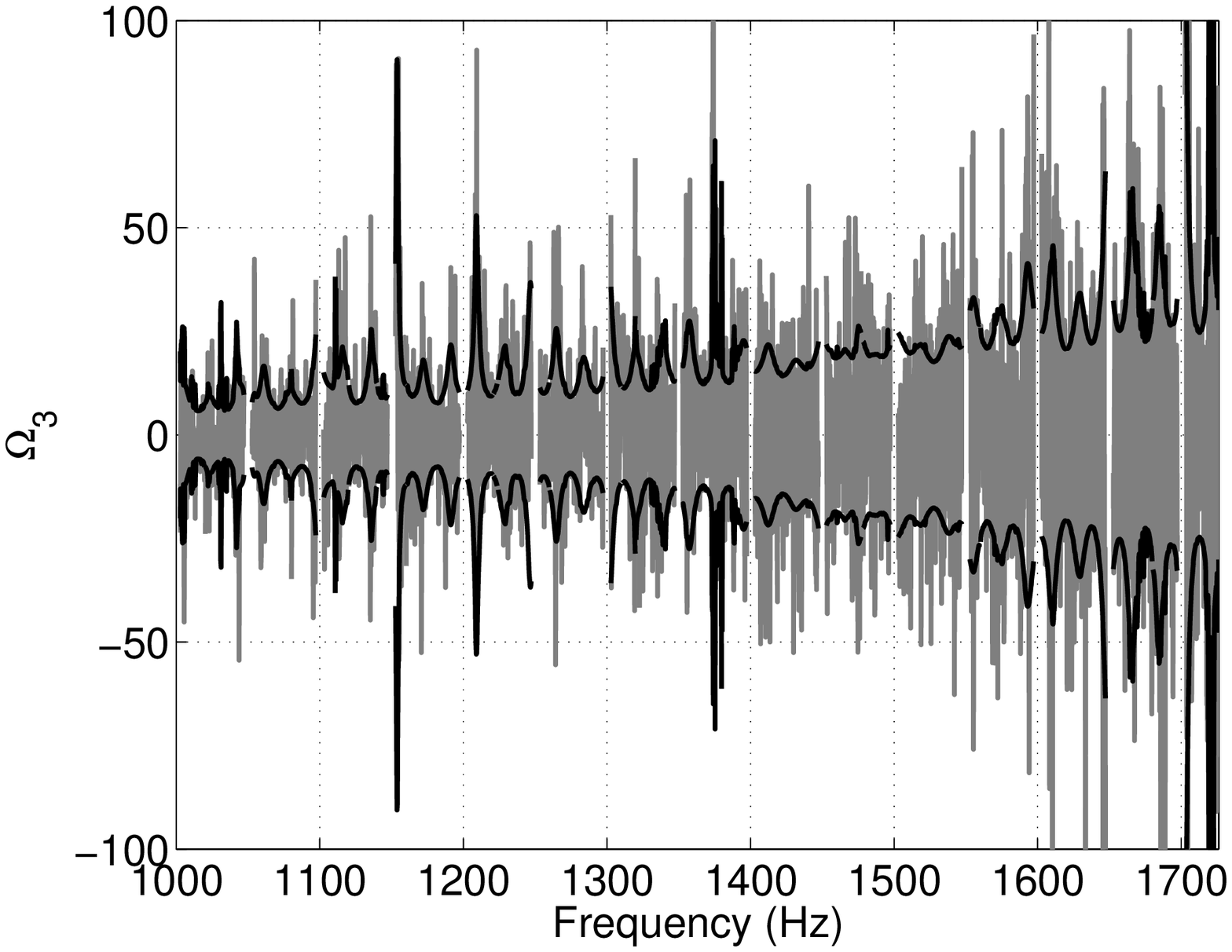, width=0.48\textwidth}
  \end{tabular}
\caption{
Integrand of Equation~\ref{eq-3} multiplied by $df=0.25~{\rm Hz}$ (gray) and the associated 1 sigma uncertainty (black).  Though energy density is a positive quantity, its estimator can be either positive or negative due to noise.  Fluctuations of the  estimator around zero are consistent with the absence of a signal.  The broadband results in Table~\ref{tbl-1} are obtained as a weighted average over each observing band following~\citep{1999PhRvD..59j2001A}.  Each spectrum includes data from all available detector pairs in 2009--2010.  LIGO and Virgo are most sensitive in the 41.5--169.25~Hz band.
}
\label{fig-1}
\end{figure*}

\emph{Results and discussion.}---Applying the previously described search techniques and data$\mbox{-}$quality cuts, we obtain results in each of four frequency bands that together span 41.5--1726~Hz and are summarized in Table~\ref{tbl-1}.  
In Figure~\ref{fig-1} we plot the frequency-dependent contributions to $\Omega_\alpha$.
We find no evidence for an isotropic gravitational$\mbox{-}$wave background and set direct upper limits on the energy density of the SGWB.

\begin{table*} 
\begin{tabular}{|c|c|c|c|c|c|}
\hline
Frequency (Hz) & $f_{\rm{ref}}$ (Hz) & $\alpha$ & $\Omega_{\alpha}$ & 95\% CL upper limit & Previous limits \\
\hline
\hline
\hline
41.5-169.25 & - & 0 & $(-1.8\pm4.3)\times10^{-6}$ & $5.6\times10^{-6}$ & $7.7\times10^{-6}$ \\
\hline
170-600 & - & 0 & $(9.6\pm4.3)\times10^{-5}$ & $1.8\times10^{-4}$ & - \\
\hline
600-1000 & 900 & 3 & $0.026\pm0.052$ & $0.14$ & $0.35$ \\
\hline
1000-1726 & 1300 & 3 & $-0.077\pm0.53$ & $1.0$ & - \\
\hline
\end{tabular}
\caption{Results of the stochastic analysis of 2009--2010 LIGO and Virgo data.  Note that the previous limits are scaled to the current best estimate of $H_0$ \citep{2013arXiv1303.5062P}. \label{tbl-1}}
\label{table1}
\end{table*}

\emph{41.5--169.25~Hz band}: We use a spectral index of $\alpha=0$, a value motivated by cosmological models, following the precedent of \citep{2009Natur.460..990A}.  Using the previous LIGO results \citep{2009Natur.460..990A} as a prior and marginalizing over detector calibration uncertainties \footnote{Marginalization over detector calibration uncertainties was performed following the method outlined in \citep{2014JPhCS.484a2027W}, using conservative amplitude calibration errors of 16.2\%, 23.0\%, and 6.0\% for H1, L1, and V1, respectively \citep{2012Kawabe}.  Phase calibration errors are negligible \citep{2012PhRvD..85l2001A}.}, we determine the 95\% confidence level (CL) upper limit to be $\Omega_0<5.6\times10^{-6}$.  This is the first result using both LIGO and Virgo data in this frequency band and it is the best direct limit on the SGWB energy density at these frequencies.  The previous S5 result in this band \citep{2009Natur.460..990A} set an upper limit of $\Omega_0<7.7\times10^{-6}$  (when scaled for the current best estimate of $H_0$ \citep{2013arXiv1303.5062P}).  The limit here is a 38\% improvement.  

\emph{600--1000~Hz band}: 
For this frequency band, we use a reference frequency of 900~Hz and a spectral index of $\alpha=3$ (an astrophysically motivated value) following \citep{2012PhRvD..85l2001A}.  After taking detector calibration uncertainties into account and using the previous LIGO/Virgo results as a prior \citep{2012PhRvD..85l2001A}, we determine the 95\% CL upper limit to be $\Omega_3<0.14$.  Previous to this result, the best direct limit in this frequency band was from the combined results of LIGO and Virgo reported in \citep{2012PhRvD..85l2001A} with $\Omega_3<0.35$ (using the present best estimate of $H_0$ \citep{2013arXiv1303.5062P}).  Our limit is a factor of 2.5 lower than this result.  This improvement comes from enhanced detector sensitivity at frequencies above 300~Hz in S6 and VSR2-3, despite a shorter observation time.

\emph{Additional frequency bands}: We report additional frequency bands spanning 170--600~Hz and 1000--1726~Hz.  For the 170--600~Hz band, we measure the 95\% CL upper limit to be $1.8\times10^{-4}$, assuming a flat prior from 0 to 1.  We find the 95\% CL upper limit to be $1.0$ for the 1000-1726~Hz band, assuming a flat prior from 0 to 10.  These are the first measurements of the SGWB in these bands.  For the 170--600~Hz band, $\Omega_{0}$ exceeds the single$\mbox{-}$sigma error bar by a factor of 2.2 which has a 10\% chance of happening due to Gaussian noise given that we analyze four independent frequency bands.

\begin{figure*} 
\centering
 \begin{tabular}{c}
    \psfig{file=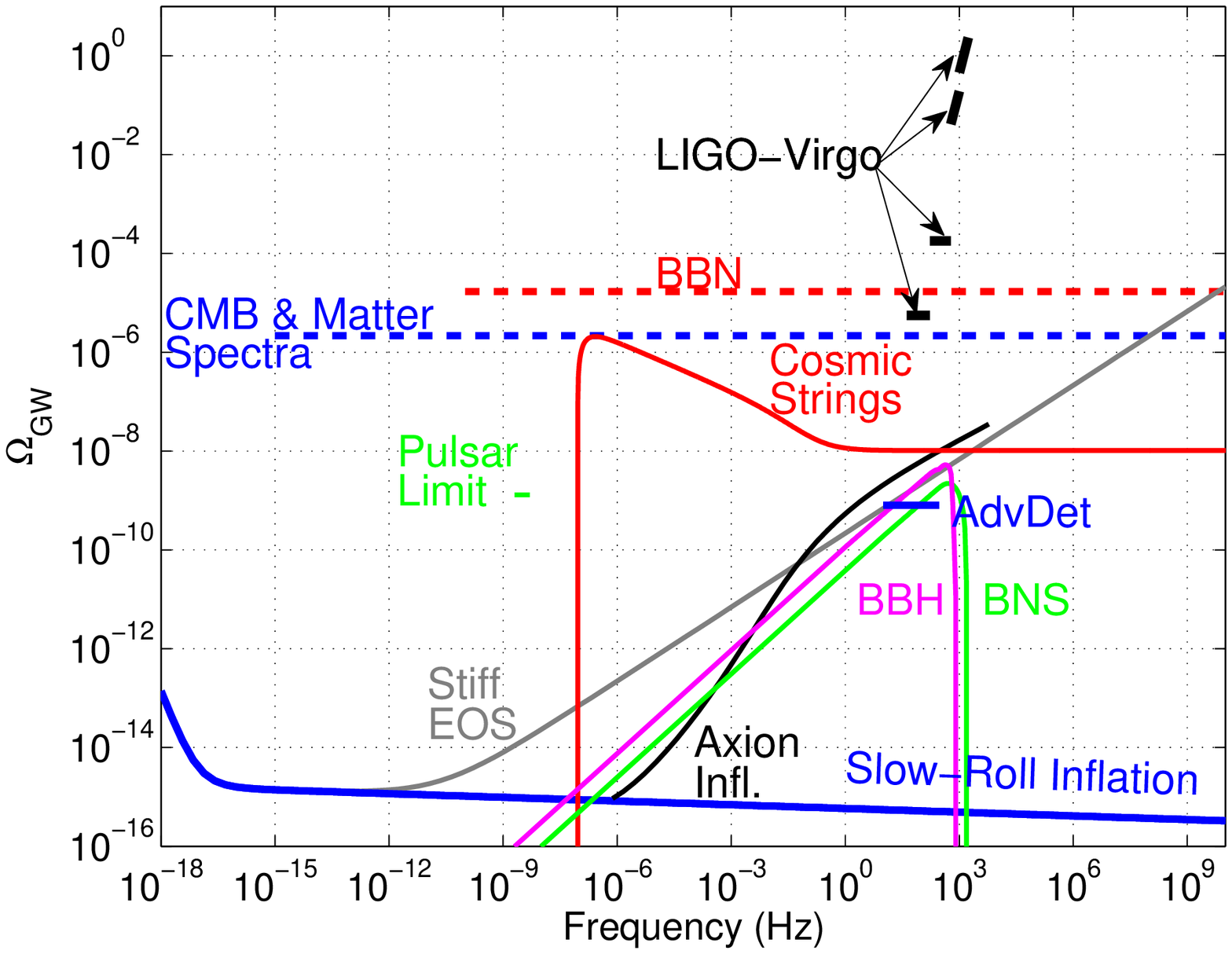, width=0.96\textwidth}
  \end{tabular}
\caption{
  Normalized GW energy density versus frequency for experimental bounds and for several SGWB models (see text for detail).  Note that the different experimental bounds shown in this figure constrain different quantities.  The LIGO-Virgo upper limits are on $\Omega_{\alpha}$ (for $\alpha = 0, 3$, see Table~\ref{tbl-1}), which are converted into bounds on $\Omega_{\rm GW}(f)$ as defined by Equation~\ref{eq-2}.  While `BBN' and `CMB \& Matter Spectra' constrain the total GW energy density in the frequency bands indicated by their respective lines, `Pulsar Limit' is on $\Omega_{\rm GW}(f)$ at the specific frequency of $f = 2.8~{\rm nHz}$.  
}
\label{fig-2}
\end{figure*}

\emph{Implications.}---Figure~\ref{fig-2} shows the upper limits from our measurement (solid black lines, denoted `LIGO-Virgo') in comparison with other bounds on the SGWB and several representative SGWB models. We include the indirect bound on the total GW energy density in the $10^{-10}-10^{10}~{\rm Hz}$ band derived from Big Bang nucleosynthesis and observations of the abundances of the lightest nuclei \citep{2000PhR...331..283M,1997rggr.conf..373A,2005APh....23..313C} (dashed red line).  We also include the similar indirect homogeneous bound from CMB and matter power spectra measurements \citep{PhysRevD.85.123002} (dashed blue line).  The bound due to millisecond pulsar timing measurements \citep{Shannon18102013} is solid green (`Pulsar Limit').  The projected sensitivity of the advanced GW detector network including Advanced LIGO \citep{2010CQGra..27h4006H}, Advanced Virgo \citep{Virgo2009}, and KAGRA \citep{2012CQGra..29l4007S} is solid blue (`AdvDet').  Recently, the BICEP2 collaboration claimed observation of B-mode polarization in the CMB and considered an interpretation where the polarization signal is largely due to tensor modes \citep{PhysRevLett.112.241101}.  A canonical slow-roll inflationary model with tensor-to-scalar ratio $r=0.2$ (the BICEP2 best fit) yields the spectrum shown by the solid blue line (`Slow-Roll Inflation'), predicting $\Omega_{\rm GW}\sim5\times10^{-16}$ in the frequency band of terrestrial GW detectors \citep{1997PhRvD..55..435T}. This signal is not within reach of the measurement described here, nor will it be within reach of the advanced detector network. Observation of this signal with GW detectors will require novel technology, possibly satellite-based \citep{bbo} or underground \citep{PhysRevD.88.122003}. 

Future measurements of this inflationary signal by GW detectors, combined with the CMB B-mode polarization measurements, will constrain the tensor spectral index $n_t$, hence constraining inflationary models \citep{2013arXiv1303.5082P}.  GW measurements hold great promise for probing the physics of inflation as well as for probing processes at the energy scales of $10^3-10^{10}~{\rm GeV}$ \citep{PhysRevD.75.043507}, well beyond those of the Large Hadron Collider.  For example, the late stages of inflation could generate boosts in the GW spectrum at high frequencies, either through a preheating resonant phase \citep{2006JCAP...04..010E,2007PhRvL..99v1301E} or via the back-reaction of fields generated by the inflaton \citep{2012PhRvD..85b3525B,2012PhRvD..85b3534C}. As shown in Figure~\ref{fig-2}, the axion-inflaton model including back-reaction (black, `Axion Infl.') could produce a GW spectrum sufficiently strong to be observed by the advanced detector network.  The evolution of the universe after inflation and before Big Bang nucleosynthesis is not well understood.  The presence of a new ``stiff" energy component at this time (with equation of state parameter $w>1/3$) could also result in a significant high-frequency boost to the GW spectrum \citep{2008PhRvD..78d3531B}.  Figure~\ref{fig-2} shows the example of $w=0.6$ (denoted `Stiff EOS'), which may also be detectable by the advanced detector network.  A cosmological background from cosmic strings (`Cosmic Strings') is potentially detectable as well \citep{2007PhRvL..98k1101S}.

It should also be noted that astrophysical GW foregrounds could mask the inflationary signal.  Figure~\ref{fig-2} shows the possible GW spectra from the stochastic superposition of all the binary neutron stars (`BNS', green) and binary black holes (`BBH', magenta) \citep{PhysRevD.85.104024}, which are too distant to be individually resolved with advanced detectors.  Realistic binary rates may lead to a detectable stochastic signal in the advanced detector network.  Other astrophysical models (including rotating neutron stars \citep{PhysRevD.87.063004,2012PhRvD..86j4007R,2011ApJ...729...59Z}, magnetars \citep{2011MNRAS.411.2549M,2006A&A...447....1R,2013PhRvD..87d2002W}, and others) may also contribute to the astrophysical foreground. Astrophysical sources are interesting in their own right.  However, foreground subtraction may be necessary to reach a slow-roll inflationary signal. Such a subtraction will require detailed understanding of the foregrounds, which in turn may require multiple detectors operating in different frequency bands to disentangle different frequency and spatial contributions \citep{PhysRevLett.109.171102}.


\emph{Conclusions.}---The results presented above include data from both LIGO and Virgo and span the frequency range of 41.5--1726~Hz.  The upper limit placed on the low frequency 41.5--169.25~Hz band is 38\% lower than previous direct measurements \cite{2009Natur.460..990A}.  For the 600--1000~Hz band, the upper limit is a factor of 2.5 lower than previous direct measurements \citep{2012PhRvD..85l2001A}.  We also place the first upper limits over the remainder of the LIGO/Virgo frequency range: 170--600~Hz and 1000--1726~Hz.  Together, these are the lowest upper limits from direct measurements of the SGWB to date.  

With Advanced LIGO and Advanced Virgo detectors on the horizon, the sensitivity of interferometers to the SGWB will improve substantially in the coming years.  This will allow us to probe astrophysical sources such as binary black holes and cosmological sources such as axion inflation.  We may also detect an unexpected source.  To reach the SGWB generated by the standard slow-roll inflationary model, however, more sensitive gravitational wave detectors will be needed, likely deploying novel technologies.

\emph{Acknowledgments.}---The authors gratefully acknowledge the support of the United States National Science Foundation for the construction and operation of the LIGO Laboratory, the Science and Technology Facilities Council of the United Kingdom, the Max-Planck-Society, and the State of Niedersachsen/Germany for support of the construction and operation of the GEO600 detector, and the Italian Istituto Nazionale di Fisica Nucleare and the French Centre National de la Recherche Scientifique for the construction and operation of the Virgo detector. The authors also gratefully acknowledge the support of the research by these agencies and by the Australian Research Council, the International Science Linkages program of the Commonwealth of Australia, the Council of Scientific and Industrial Research of India, the Istituto Nazionale di Fisica Nucleare of Italy, the Spanish Ministerio de Econom{\'i}a y Competitividad, the Conselleria d'Economia Hisenda i Innovaci\'{o} of the Govern de les Illes Balears, the Foundation for Fundamental Research on Matter supported by the Netherlands Organisation for Scientific Research, the Polish Ministry of Science and Higher Education, the FOCUS Programme of Foundation for Polish Science, the Royal Society, the Scottish Funding Council, the Scottish Universities Physics Alliance, the National Aeronautics and Space Administration, the National Research Foundation of Korea, Industry Canada and the Province of Ontario through the Ministry of Economic Development and Innovation, the National Science and Engineering Research Council Canada, the Carnegie Trust, the Leverhulme Trust, the David and Lucile Packard Foundation, the Research Corporation, and the Alfred P. Sloan Foundation.

\bibliography{iso_bibliography_v2}

\end{document}